\documentclass[useAMS,usenatbib]{mn2e}
\usepackage{epsfig}
\usepackage{color}
\usepackage{comment}

\def\beqra{\begin{eqnarray}}
\def\eeqra{\end{eqnarray}}
\def\beq{\begin{equation}}
\def\eeq{\end{equation}}

\def\ltsima{$\; \buildrel < \over \sim \;$}
\def\simlt{\lower.5ex\hbox{\ltsima}}
\def\gtsima{$\; \buildrel > \over \sim \;$}
\def\simgt{\lower.5ex\hbox{\gtsima}}
\title[Spatial \& velocity bias of density peaks and protohaloes]{The spatial and velocity bias of linear density peaks and protohaloes in the $\Lambda$ cold dark matter cosmology}
\author[A. Elia, A. Ludlow and C. Porciani]{Anna Elia\thanks{E-mail:
elia@astro.uni-bonn.de}, Aaron D.
Ludlow and Cristiano Porciani \\
Argelander Institut f\"ur Astronomie der Universit\"at Bonn, Auf dem
H\"ugel 71, D-53121 Bonn, Germany}
\begin{document}
\pagerange{\pageref{firstpage}--\pageref{lastpage}} \pubyear{2011}
\maketitle

\label{firstpage}

\begin{abstract}
We use high resolution N-body simulations to investigate the Lagrangian bias of cold dark matter haloes within the $\Lambda$ cold dark matter cosmology. Our analysis
focuses on `protohaloes', which we identify in the simulation initial conditions with the subsets of particles belonging to individual redshift-zero haloes.
We then calculate the number-density and velocity-divergence fields of protohaloes and estimate their auto spectral densities. We also measure the corresponding cross spectral densities with the linear matter distribution. We use our results to test a Lagrangian-bias model presented by Desjacques and Sheth which is based on the assumption that haloes form out of local density maxima of a specific height. Our comparison validates the predicted functional form for the scale-dependence of the bias for both the density and velocity fields. We also show that the bias coefficients are accurately predicted for the velocity divergence. On the contrary, the theoretical values for the density
bias parameters do not accurately match the numerical results as a function of halo mass. This is likely due to the simplistic
assumptions that relate virialized haloes to density peaks of a given height in the model. We also detect appreciable stochasticity for the Lagrangian density bias, even on very large scales. These are not included in the model at leading order but correspond to higher order corrections. 
\end{abstract}


\section{Introduction}
Galaxy redshift surveys are powerful probes of cosmology.
The main observables able to constrain cosmological parameters are the overall shape of the galaxy power spectrum at wavenumbers $k<0.1\, \mathrm{Mpc}^{-1}$ and the baryonic acoustic oscillations within it. These are treated as proxies for the matter power spectrum for which we can make robust theoretical predictions.
Galaxies, however, are biased tracers of the cosmic mass distribution and many features appearing in their power spectrum depend on how a specific observational
sample was selected. To reconstruct the matter power spectrum we thus need an accurate bias model whose free coefficients should
be used as nuisance parameters and marginalized over. In the era of precision cosmology, where measurements of the matter power spectrum with per cent accuracy are required, this task is particularly demanding. 

Bias models can be divided into two broad classes.
Eulerian biasing schemes relate the galaxy density contrast, $\delta_{\rm g}({\bf x},t)$, 
to the matter density distribution, $\delta$, evaluated at the same time $t$ (but not necessarily at the same spatial location).
After smoothing the fields on large scales, so that $|\delta|$ is typically much smaller than unity, one can write \citep{b27}
\beqra
\label{nonloceul}
\delta_{\rm g}({\bf x})&=&B_0+\int d^3x_1\,B_1({\bf x}-{\bf x_1})\,\delta({\bf x_1})+\\
&+&\frac{1}{2}
\int d^3x_1\,d^3x_2\,B_2({\bf x}-{\bf x_1},{\bf x}-{\bf x}_2)\,\delta({\bf x_1})\,\delta({\bf x_2})+ \nonumber\\
&+& \dots \,, \nonumber
\eeqra
where all fields are evaluated at the same time $t$ and the details of the bias model are specified by the kernel functions, $B_{\rm i}$.
If one further assumes that biasing is local (i.e. that all kernels can be written as products of Dirac delta distributions), this reduces to 
\beq
 \delta_{\rm g}({\bf x})=b_0+b_1\,\delta({\bf x})+\frac{b_2}{2}\,\delta^2({\bf x})+\dots \,,
\label{loceul}
\eeq
where now the bias coefficients $b_{\rm i}$ are real numbers.

In Lagrangian bias models, on the other hand, one considers the regions
in the initial conditions that will collapse to form galaxies (or their hosting dark-matter
haloes) at time $t$ 
and writes their density contrast, $\delta^{\rm L}_{\rm g}({\bf q})$,
in terms of the linear density contrast, $\delta_0({\bf q})$.
Large-scale expansions analogous to eqs. (\ref{nonloceul}) and (\ref{loceul}) can also be written in this case.
As a second step,
one must determine the final position, ${\bf x}({\bf q},t)$, of a fluid
element initially located at ${\bf q}$,
and compute $\delta_{\rm g}({\bf x}({\bf q},t),t)$ 
out of $\delta^{\rm L}_{\rm g}({\bf q})$.
This Lagrangian-to-Eulerian mapping (LEM) accounts for gravitationally induced 
motions that determine the final position of the objects.

Local Eulerian and Lagrangian bias schemes are not equivalent; they generate a different
shape for the galaxy bispectrum \citep{b28} and are not compatible within the framework
of perturbation theory \citep{b29}.
In fact,
\citet{b30} have shown that a local Lagrangian biasing scheme 
generates a non-linear, non-local and stochastic bias in Eulerian space.
On the other hand, non-local Eulerian and Lagrangian schemes are equivalent and can be seen as different mathematical representations
of the same physical process \citep{b29}.

Due to its simplicity, the local Eulerian model is by far the most
widely used in practical applications, such as perturbative calculations.
However, it is purely phenomenological and does not have a 
strong theoretical motivation. Detailed comparison with numerical simulations has also evidenced 
its limited validity (e.g. \citealt{b31}, \citealt{b32}).
Physical models of bias are generally given in the Lagrangian framework as conditions on galaxy (halo) formation are more easily imposed onto the linear 
density field using some model for the collapse of density perturbations.
\citet[][hereafter MW]{b22} used a Press-Schechter-like argument \citep{b34} to compute the bias coefficients of a local Lagrangian scheme as a function of halo mass. The same authors also showed how these parameters 
can be combined to calculate the bias coefficients of a local Eulerian scheme 
assuming that large-scale density perturbations follow the spherical collapse 
model. The effect of non-linear shear on the LEM was discussed
by \citet{b30} using the Zel'dovich approximation \citep{b33}.
For halo masses $M>M_*(z)$ (where $M_*(z)$ is the characteristic mass for collapse at redshift $z$) the MW formula for $b_1$ is in 
good agreement with the predictions of N-body simulations \citep{b26}. 
At lower masses, however, the agreement rapidly deteriorates \citep{b35}.
\citet{b23} and \citet{b36} showed that the discrepancy
between the N-body simulations and the analytical predictions is already
present in Lagrangian space and should thus be attributed to the
limitations of the Press-Schechter formalism rather than to the approximated
treatment of the LEM.

The Lagrangian bias emerging from the extended Press-Schechter model 
\citep{b38} was first derived by
\citet{b37}, rediscussed in \citet{b39}
and tested against simulations by \citet{b40}.
This approach follows correlated trajectories of $\delta_0$ at different
Lagrangian locations as a function of the smoothing scale and looks
for correlations in the first-crossing scales of a density threshold.

According to the peak-background-split argument (\citealt{b2}; \citealt{b21}),
long-wavelength density fluctuations modulate 
halo formation by modifying the collapse time of localized short-wavelength
perturbations. This makes it possible to generalise the calculation of the Lagrangian MW bias 
coefficients to any model for the halo mass function
(\citealt{b41}, \citealt{b43}, \citealt{b44})
and to improve the agreement with N-body simulations
(\citealt{b45}, \citealt{b42}, \citealt{b46}, \citealt{b6}).

\citet{b15} explained the strong clustering of 
Abell clusters by assuming that they originate from the regions above a density threshold in the (suitably smoothed) 
linear density field. 
Following this line of reasoning, it is common to assume that dark-matter haloes form out of linear density peaks, as an alternative to
the Press-Schechter approach.
Tests against N-body simulations have shown that this is a well justified assumption,
especially for massive haloes (\citealt{b8}, see also \citealt{b51}).
The statistical properties of the local maxima in a Gaussian random field  
have been extensively studied by \citet{b2} (see also \citealt{b16} and \citealt{b17}).
\citet{b26} introduced peaks theory in the MW formalism,
while \citet{b24} evaluated the level of stochasticity in the Lagrangian clustering of density extrema.

Recently, \citet{b7} and \citet[][hereafter DS]{b1}
showed that the correlation of density peaks in real and redshift space
can be interpreted in terms of a simple Lagrangian biasing scheme.
Due to the peak constraint, the effective Lagrangian peak density at a given point 
not only depends on the local value of the mass density but also on its Laplacian.
The first-order peak bias depends on the mass and height of the peaks, and on the matter power spectrum.
For high peaks, this reduces to the results by \citet{b24}.
Moreover, although peaks move with the matter at their positions, DS infer the existence
of a statistical velocity bias due to the fact that
local maxima can only exist at special locations.
This also leads to a bias between the linear velocity spectra
of peaks and matter which is predictable in quantitative terms.

In spite of the fact that the DS model provides the initial conditions for 
sophisticated models of the (Eulerian) halo distribution
where the LEM is based either on 
resummed perturbation theory
\citep{b48} or on the Zel'dovich approximation \citep{b47},
its predictions for the Lagrangian clustering and velocities of the regions 
that will form collapsed structures  
have never been thoroughly tested against numerical simulations.
This paper provides such a test, which is necessary if we are to use advanced bias models 
to extract useful information on the cosmological parameters 
through a comparison with observations.

The structure of the paper is as follows. In Section \ref{mod} we review the bias
model first presented in DS. The details of our N-body simulations and
analysis techniques are outlined in Section \ref{numerics}, and a comparison between
the model and numerical results is presented in Section \ref{tre}. Finally, we
summarize our main results in Section \ref{sum}.

\section{The DS model}
\label{mod}

In this section we summarize, for completeness, the peaks model described in Section 2 of DS. In 
order to do so, we first introduce and define some quantities relevant for peak statistics.

The spectral moments of the matter power spectrum are defined as
\beq
\sigma_{\rm n}^2 (R_{\rm s},z) = \frac{1}{2 \pi^2} \int_0^{\infty} \mathrm{d}k\,k^{2(n+1)}\,P(k,z)\,W(k,R_{\rm s})^2\,,
\label{a}
\eeq
where $P(k,z)$ is the linear matter power spectrum at redshift $z$, and $W(k,R_{\rm s})$ is a smoothing
kernel of characteristic length $R_{\rm s}$. In terms of the moments we define the spectral 
parameters:
\beq
\gamma_{\rm n} \equiv \frac{\sigma_{\rm n}^2}{\sigma_{\mathrm{n-1}}\sigma_{\mathrm{n+1}}}\,.
\label{b}
\eeq
There are two common choices for $W(k,R_{\rm s})$: the Gaussian filter
\beq
W_{\rm G}(k,R_{\rm s}) = \mathrm{e}^{-\frac{k^2 R_{\rm s}^2}{2}}\,,
\label{c}
\eeq
and the top-hat filter
\beq
W_{\rm TH}(k,R_{\rm s}) = \frac{3 [\sin(k R_{\rm s})-k R_{\rm s} \cos(k R_{\rm s})]}{k^3 R_{\rm s}^3}\,.
\label{d}
\eeq
The mass contained within the comoving length $R_{\rm s}$ in these cases is
\beq
M_{\rm G}(R_{\rm s}) = (2\pi)^{3/2} \bar{\rho} R_{\rm s}^3 \textrm{, and  } \qquad M_{\rm TH}(R_{\rm s}) = \frac{4\pi}{3} \bar{\rho} R_{\rm s}^3\,,
\label{e}
\eeq
where $\bar{\rho}$ is the mean matter density of the Universe.

We will often characterize peaks in terms of their dimensionless peak height, $\nu$, defined
\beq
\nu(R_{\rm s},z_{\rm c}) = \frac{\delta_{p}}{\sigma_0(R_{\rm s},z_{\rm c})}\,.
\label{h}
\eeq
Here $\delta_{\rm p}$ is the smoothed overdensity at the peak location linearly extrapolated to $z=0$, and $\sigma_0(R_{\rm s},z_{\rm c})$ is the
linear rms mass fluctuation in spheres of radius $R_{\rm s}$. DS linked density maxima of height $\nu(R_{\rm s},z)$ to dark matter
haloes of mass $M_{\rm s}$ collapsing at redshift $z_{\rm c}$, assuming that $\delta_{\rm p}$ coincides with the threshold for collapse, $\delta_{\rm c}$.

\citet{b2} and DS computed the cross-correlation between peaks and the underlying density field
which also corresponds to the average density profile around density maxima.
Similarly, \citet{b7} evaluated the leading order expressions (on large spatial separations) for the peak auto-correlation function  
and the line-of-sight mean streaming for pairs of discrete local maxima of height $\nu$.
\citet{b7} and DS showed that the full expression for the cross-correlation and the large-scale asymptotic of the auto-correlation  
function are consistent with -- and thus can be thought of as arising from -- an effective bias relation.
In the DS model, the number density and the velocity of peaks, $\delta n_{\rm pk}$ and ${\bf v}_{\rm pk}$, are related to the dark matter density 
contrast and velocity fields, linearly extrapolated to $z=0$, via:
\beq
\delta n_{\rm pk} ({\bf x}|z_{\rm c}) = b_{\nu}\delta_{\rm S}({\bf x})-b_{\zeta} \nabla^2\delta_{\rm S}({\bf x}) ,
\label{ia}
\eeq
and
\beq
{\bf v}_{\rm pk} ({\bf x}|z_{\rm c}) = {\bf v}_{\rm S} ({\bf x}) - \frac{\sigma_0^2}{\sigma_1^2} \nabla \delta_{\rm S}({\bf x}) \,.
\label{ib}
\eeq
The subscript ``S'' indicates that the fields are smoothed on the scale $R_{\rm s}$, and the bias parameters,
$b_{\nu}$ and $b_{\zeta}$, are given by
\beq
b_{\nu} = \frac{1}{\sigma_0} \left(\frac{\nu-\gamma_1 \bar{u}}{1-\gamma_1^2} \right)\,, 
\label{ja}
\eeq
and
\beq
b_{\zeta} = \frac{1}{\sigma_2} \left(\frac{\bar{u}-\gamma_1 \nu}{1-\gamma_1^2} \right)\,.
\label{jb}
\eeq
Here $\bar{u}$ is the mean curvature of the peaks, which can be approximated by \citep{b2}
\beq
\bar{u}=\gamma_1 \nu + \frac{3(1-\gamma_1^2)+(1.216-0.9\gamma_1^4) \exp \left[-\frac{\gamma_1}{2} \left(\frac{\gamma_1 \nu}{2} \right)^2 \right] }{\left[3(1-\gamma_1^2)+0.45+\left(\frac{\gamma_1 \nu}{2} \right)^2 \right]^\frac{1}{2} + \left(\frac{\gamma_1 \nu}{2} \right) } \,.
\label{k}
\eeq
Note that $b_{\nu}$ coincides with the peak bias factor found by \citet{b2} after neglecting
the derivatives of the density correlation function.

Since, by definition, the gradient of the density field vanishes at peak
locations, eq. (\ref{ib}) suggests that the peak and dark matter velocity fields must be coincident there. By construction,
{\em peaks move with the dark matter flow}, yet the spatial bias induces a statistical velocity bias.
We will consider the scaled velocity divergence, $\theta({\bf{x}})=\nabla \times {\bf v}({\bf{x}})/(aHf)$,
rather than the velocity field. Here $a$ is the scale factor, $H$ the Hubble parameter, $f=d\ln D/d\ln a$, with
$D$ the linear growth factor. In these units, both $\theta({\bf{x}})$ and $\delta({\bf{x}})$ are 
dimensionless quantities. With these changes, eqs. (\ref{ia}) and (\ref{ib}) can be
rewritten in Fourier space as
\beq
\delta n_{\rm pk} ({\bf k}) = (b_{\nu} +b_{\zeta} k^2)\, \delta({\bf k})\, W(k,R_{\rm s}) 
\label{la}
\eeq
and
\beq
\theta_{\rm pk} ({\bf k}) = \left(1 - b_{\sigma} k^2 \right) \theta ({\bf k})\, W(k,R_{\rm s}) = b_{\rm vel}(k)\theta ({\bf k})\,,
\label{lb}
\eeq
where we have defined 
\beq
b_{\sigma} = \frac{\sigma_0^2}{\sigma_1^2}\,.
\label{m}
\eeq
In the limit of high peaks ($\nu \gg 1$) it can be shown that the bias parameters obey the following
asymptotic relations: $b_{\nu} \to \nu/\sigma_0$  and $b_{\zeta} \to 0$. This implies that the highest 
peaks are linearly biased tracers of the underlying matter field. This is consistent with the predictions of the peak-background split \citep{b26}
and DS showed that, indeed, $b_{\nu}$ is the appropriate generalization of the constant, large-scale bias
for low $\nu$. Unlike the density bias factors, $b_{\sigma}$ does not depend on $\nu$.

In order to test this model against N-body simulations, we will make use of the cross-spectra between the 
peak and dark matter densities and velocities (denoted as $P_{\mathrm{mp}}$ and $\mathcal{P}_{\mathrm{mp}}$, respectively)
and of the corresponding peak auto-spectra ($P_{\mathrm{p}}$ and $\mathcal{P}_{\mathrm{p}}$).
From eqs. (\ref{la}) and (\ref{lb}) we obtain 
\beqra
P_{\mathrm{mp}}(k) &=& (b_{\nu} +b_{\zeta} k^2)\,P(k)\,W(k,R_{\rm s}) \,, \nonumber \\
P_{\mathrm{p}}(k) &\simeq& (b_{\nu} +b_{\zeta} k^2)^2\,P(k)\,W^2(k,R_{\rm s}) \,, 
\label{n}
\eeqra
\beqra
\mathcal{P}_{\mathrm{mp}}(k) &\simeq& (1-b_{\sigma} k^2)\,\mathcal{P}(k)\,W(k,R_{\rm s}) \,, \nonumber \\
\mathcal{P}_{\mathrm{p}}(k) &\simeq& (1-b_{\sigma} k^2)^2\,\mathcal{P}(k)\,W^2(k,R_{\rm s}) \,,
\label{nb}
\eeqra 
where $P(k)$ and $\mathcal{P}(k)$ are the matter density and velocity divergence auto-spectra, respectively.
We remind the reader that the expression for $P_{\rm p}(k)$ is only valid to first order in $P(k)$ as $k \to 0$, and that higher order corrections should be included to improve its accuracy (see e.g. \citealt{b47}). On the contrary, the expression for the cross-spectrum $P_{\rm mp}$ is  
exact, as shown in \citet{b2} and in the Appendix A of DS.
Eq. (\ref{n}) has the same functional form as eq. (57) in \citet{b24} who studied the clustering of density extrema.
The DS coefficients $b_{\nu}$ and $b_{\zeta}$ match those in \citet{b24} only in the limit $\nu \gg 1$, for which nearly all  
extrema are density maxima.

\section{Numerical Issues}
\label{numerics}

In this section we provide a brief description of the main numerical issues relevant for this work. This includes a brief
description of our numerical simulations in Section \ref{num}, our main analysis techniques in Section \ref{analysis}, 
and a characterization of halo collapse barriers in Section \ref{hei}.

\subsection{N-body simulations}
\label{num}

Our analysis focuses on two high-resolution N-body simulations of structure formation in the standard LCDM cosmology. The cosmological parameters for our runs were chosen to be consistent with the fifth-year WMAP data release \citep{b4}.
These are $h=0.701$, $\sigma_8=0.817$, $n_{\rm s}=0.96$, $\Omega_{\rm m}=0.279$, $\Omega_{\rm b}=0.0462$ and $\Omega_{\Lambda}=1-\Omega_{\rm m}=0.721$.
Each simulation was run with a lean version of the Tree-PM code \textsc{Gadget}-2 \citep{b3}, and followed the dark matter using 
$1024^3$ collisionless particles.
One simulation had a box side-length of $L_{\rm box}=1200\, h^{-1}$ Mpc and a particle mass
of $m_{\rm part}=1.246\cdot 10^{11} h^{-1} M_{\odot}$; the other used $L_{\rm box}=150\, h^{-1}$ Mpc and had $m_{\rm part}=2.433\cdot 10^{8} h^{-1} M_{\odot}$. The initial redshifts of the simulations were $z_{\rm in}=50$ and
$z_{\rm in}=70$ for the larger and smaller box, respectively. Using these simulations we are able to probe a wide range of halo masses, spanning $8\cdot 10^{10} h^{-1} M_{\odot} < M_{\rm h} < 10^{14} h^{-1} M_{\odot} $.
These simulations were first studied in \citet{b6}, and later by
\citet{b8}, and we refer the reader to those papers for further
details.

Haloes were identified at $z=0$ using a friends-of-friends (FOF) algorithm \citep{b25}
with a linking length of 0.2 times the mean interparticle distance. 
Protohaloes were identified by tracing backward to the initial redshift of all subsets of the particles belonging to
$z=0$ FOF haloes. We use the centre of mass of each protohalo as a proxy for its spatial location; 
the mass-weighted linear velocity provides an estimate of the protohalo's motion.
As a test of the sensitivity of our results to the adopted halo finder, we also generated a spherical overdensity (SO) halo 
catalogue with an overdensity threshold of 200 times the critical density, $\rho_{\rm c}$. For a fixed halo mass, the two halo-finders
produce results consistent within $10$ per cent (in terms of all the bias coefficients) and so, in what follows, we will
focus on results obtained for the FOF haloes, and only consider those
containing at least 100 particles.

Haloes in each simulation are split into four separate mass bins in order to preserve their
peculiar clustering properties. These bins are referred to as 1S to 4S for the small 
box, and 1L to 4L for the large one. To asses the impact of shot noise in the analysis
of our small-box simulation we consider an additional mass bin, labeled bin0S, which
includes all haloes with $N\geq 100$. The mass ranges and total number of haloes in each
bin are given in Table \ref{tab:1}. We note that bins with label `S' refer to masses $M<M_{*}$ 
for which dark-matter haloes are not expected to be in one-to-one correspondence with linear peaks \citep{b8}.

\subsection{Analysis}
\label{analysis}

We construct protohalo density and momentum fields 
using cloud-in-cell grid assignment on a $512^3$ mesh. 
Velocity fields are obtained by taking the ratio of the momentum and density
fields, as described in \citet{b5}. In the case of haloes, these distributions are smoothed to 
preclude the existence of empty cells;
the smoothing scales used are $R_{\rm f} = 7 h^{-1}$ Mpc for the large box and $R_{\rm f} = 1.8 h^{-1}$ Mpc for the small one. 
These values are chosen in order to mask the effects of the grid, but we have explicitly verified that our results are 
not significantly affected by them. All power spectra have been computed 
using a fast Fourier transform technique. 

The discreteness of dark-matter particles and haloes gives rise to a shot-noise
component in the spectra. For the density fields, the estimated power spectrum, $\hat{P}$, includes a shot-noise term which 
is inversely proportional to the number density of objects $\bar{n}$ (assuming Poisson sampling):
\beq
\hat{P}=P_{\mathrm{true}} + \frac{1}{\bar{n}}\,.
\label{oo}
\eeq
Shot noise is therefore negligible for the matter spectra but
may be significant for that of the haloes. The issue is more severe in Lagrangian space, because fluctuations in
the initial conditions are small. We will consider two alternative estimates of the protohalo bias; one
is determined from the shot-noise corrected auto-spectrum,
\beq
b(k) \equiv \sqrt{\frac{P_{\mathrm{h}}(k)}{P(k)}}\,,
\label{p}
\eeq
and the other from the cross-spectrum,
\beq
b_{\mathrm{eff}}(k) \equiv \frac{P_{\mathrm{mh}}(k)}{P(k)}\,.
\label{q}
\eeq 
Here the subscript ``h'' indicates the halo fields, and ``m'' the matter field (the analogous fields
for peaks are indicated with the subscript ``p''.) The relation between the two is
\beq
b_{\mathrm{eff}}(k) = b(k) \cdot r(k)\,,
\label{r}
\eeq
where $r$ is the linear correlation coefficient, defined as
\beq
r(k)=\frac{P_{\mathrm{mh}}(k)}{\sqrt{P(k) P_{\mathrm{h}}(k)}}\,.
\label{rr}
\eeq
These relations tell us that the two definitions of the bias are equivalent \emph{only if} the bias is purely deterministic in Fourier space, i.e. $r=~1$.
At leading order, in the model presented by DS, $b_{\mathrm{eff}}(k)= b_{\nu}+b_{\zeta}\,k^2$ and $b(k)=|b_{\mathrm{eff}}(k)|$ (neglecting the filter function). Any stochasticity (represented by the higher order corrections in $P_{\rm p}$) will degrade the correlation, yielding different estimates for the bias.
Because of this, we will consider both the cross- and auto-spectra to check for a potential stochastic element of the bias. \footnote{Although the effective first-order bias model by DS is deterministic in Fourier space, it is stochastic in configuration space (\citealt{b24}, \citealt{b1}).}

As for the density, we can define two estimates for the velocity bias, that we denote $b_{\theta}$ and $b_{\theta\mathrm{eff}}$,
with a correlation coefficient $r_{\theta}$.
There is no conclusive way to subtract the shot noise for the velocity divergence spectrum.
Hence we used our simulations to gain some insight into this issue. We introduced an artificial shot noise in the matter spectrum $\mathcal{P}$
by randomly drawing a fraction of the particles and found that $\mathcal{P}_{\rm shot}(k) \simeq A \cdot k^2$ asymptotically for large $k$. Therefore we fitted the amplitude factor $A$ to obtain shot-noise corrected power spectra. Note however that other terms could be important at smaller wavenumbers; in this work we only consider data for which  $\mathcal{P}_{\rm shot}(k) < 0.1 \hat{\mathcal{P}}$. \\

The DS model describes the biasing of linear density peaks that are expected to form haloes of a given mass at
a specified redshift according to some collapse model (which determines the value for $\delta_{\rm c}$). However, we analyse the density and velocity fields for the actual FOF and SO protohaloes. These are the quantities of physical interest for studying galaxy clustering in terms of halo occupation models (e.g. \citealt{b49}).
\citet{b8} showed that the vast majority of haloes in our N-body simulations can be unambiguously associated
with linear peaks in the initial conditions when smoothed on the mass scale of the halo.
For example, $\simgt 70$ per cent of all haloes can be matched with similar-mass peaks in 
$\delta_{\rm S}(\mathbf{x})$, and $\simgt 90$ per cent for haloes with $M \simgt 10^{14} h^{-1} M_{\odot}$. 
Note, however, that the correspondence between the DS peaks and protohaloes will not be perfect.\footnote{The measured abundance of haloes in the simulations is of the same order of magnitude as (albeit a bit higher than) the number of peaks with $\nu > 1.68/\sigma_0$ computed as in \citet{b2}.} On a given mass scale, some peaks with overdensities at the collapse threshold will evolve into substructures contained within larger virialized haloes (the so-called cloud-in-cloud
problem), or to haloes of significantly different mass. On top of this, numerical simulations have shown that the collapse threshold $\delta_{\rm c}$ for haloes of a given mass and redshift has a broad probabilistic distribution rather than a fixed value (\citealt{b50}; \citealt{b10}) and possibly also depends on the form of the adopted smoothing kernel. We consider some of these issues in the following section. 

\begin{figure}
\includegraphics[width = 3.0in,keepaspectratio=true]{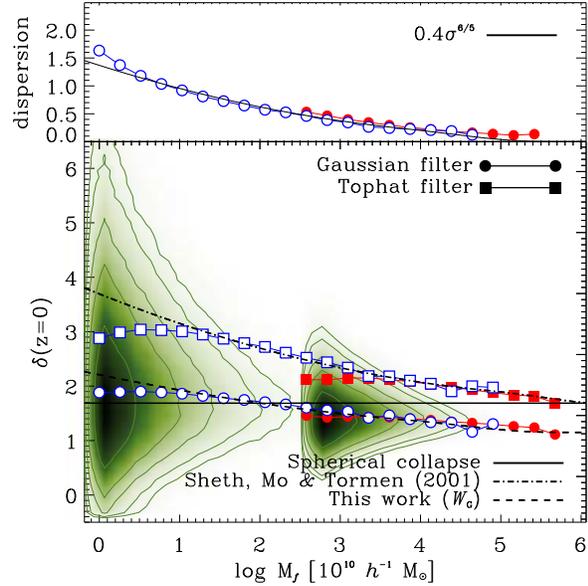}
\caption{
Mass-dependence of the linear overdensities measured at protohalo centres
in the initial conditions of our simulations. The shaded regions
and contours show the full halo distribution after smoothing with a
Gaussian filter; connected circles highlight the median trend. All values
of $\delta_h$ have been linearly extrapolated to $z=0$. Open and filled points
correspond to our $150\, h^{-1}$ Mpc and $1200\, h^{-1}$ Mpc boxes, respectively. For
comparison, we also show, using boxes, the median trends obtained after
smoothing with a top-hat kernel. In both cases, the median trends are well
described by eq. (\ref{o}): the dot-dashed curve shows the SMT result (note that
this is not a fit to our simulation data), and the dashed curve shows the
result for slightly different values of the free parameters: $\delta_{*}=1.15$, $\alpha=0.37$ and $\beta=0.515$. The top panel shows the mass dependence of the
scatter about the median trend for the Gaussian-filtered case, which is
well approximated by a simple power-law, $\Sigma = 0.4 \ \sigma_0^{6/5}$.}
\label{fig1}
\end{figure}

\subsection{Barrier heights for top-hat and Gaussian filters}
\label{hei}

The peak model described in Section ~\ref{mod} requires well-defined spectral moments.
However, due to its sharp boundary in real space, the top-hat filter decays very slowly in Fourier space so that
the integral defining $\sigma_2^2$ is divergent in a LCDM model.
Since eqs. (\ref{ja}), (\ref{jb}) and (\ref{k}) depend on $\sigma_2$, DS instead adopted a Gaussian filter and assumed $\delta_{\rm c}=\delta_{\mathrm{sc}}=1.68$. Here, $\delta_{\rm c}$ corresponds to the critical density for the collapse
of a spherical top-hat perturbation in an otherwise unperturbed
EdS universe, and this does not necessarily apply to peaks in a smoothed
Gaussian random field.
Another issue is that the validity of the simple spherical collapse model is questionable, at best; the probability of a 
protohalo or a peak being spherical is null, since it would require the three eigenvalues of the tidal tensor 
being equal. \citet{b9} showed that the barrier height
in the more general ellipsoidal collapse model \citep{b14} can be approximated by
\beq
\delta_{\mathrm{ec}}(M,z_{\rm c})=\delta_*\, \left\{ 1+ \alpha \cdot \left[\frac{\sigma_0^2(R_{\rm s}(M),z_{\rm c})}{\delta_*^2} \right]^\beta \right\}\,, 
\label{o}
\eeq
where $\delta_*=\delta_{\rm sc}$ is taken from the spherical collapse model, and $\alpha=0.47$ and $\beta=0.615$ are determined 
from fits to the model results. The presence of the dispersion,
$\sigma_0$, in eq. ~(\ref{o}) results in a mass-dependent barrier height: lower mass haloes require, on average,
higher overdensities for collapse since they must hold themselves together against larger tidal forces.
It should be emphasized that eq. (\ref{o}) describes the {\em mean} barrier height; the scatter about the mean can be approximated by
$\Sigma=0.3 \ \sigma_0$ \citep{b10}. These values are valid only for the top-hat filter. 

What is the appropriate barrier height corresponding to a Gaussian filter? In Figure \ref{fig1}
we show the linear overdensities measured at protohalo centres of mass after smoothing with a Gaussian filter
on the halo mass scale. The shaded regions show the halo data, and connected circles the medians of the
distribution. Open points correspond to results from our $150\, h^{-1}$ Mpc box
simulation, and solid points to our $1200\, h^{-1}$ Mpc box run. Squares show the
median $\delta_{\rm c}(M)$ for the same sample of haloes, but after smoothing
the linear density field with a top-hat filter instead (note the good agreement with the result of \citet{b9}, shown as a dot-dashed line in Figure \ref{fig1}). All values have been linearly extrapolated to $z=0$.

\begin{table*}
\begin{tabular}{|c|c|c|c|c|c|c|c|c|c|}
\hline
Bin & Mass range  & $\#$ haloes & $\bar{M}$  & \multicolumn{2}{|c|}{$b_{\nu}$} & \multicolumn{2}{|c|}{$b_{\zeta}$} & $b_{\sigma}$ & $R_{\rm s}$ \\ \hline
 & ($10^{12} h^{-1}\,M_{\odot}$) & & ($10^{12} h^{-1}\,M_{\odot}$) & \multicolumn{2}{|c|}{} & \multicolumn{2}{|c|}{($h^{-2}\,\mathrm{Mpc}^2$)} & ($h^{-2}\,\mathrm{Mpc}^2$) & ($h^{-1}$ Mpc) \\ \hline
& & & & SG & EG & SG & EG & & \\ \hline
0S &  $0.08-0.8$ & $106746$ & $0.2986$ & -0.31 & -0.23 & 0.63 & 0.61 & 0.92 & 0.62\\
1S &  $0.08-0.1$ & $21990$ & $0.08961$ & -0.30 & -0.21 & 0.26 & 0.25 & 0.45 & 0.42\\
2S &  $0.1-0.15$ & $30610$ & $0.1235$ & -0.30 & -0.22 & 0.33 & 0.32 & 0.55 & 0.46\\ 
3S &  $0.15-0.3$ & $32322$ & $0.2165$ & -0.31 & -0.23 & 0.49 & 0.48 & 0.76 & 0.56\\ 
4S &  $0.3-0.8$ & $21824$ & $0.4934$ & -0.31 & -0.24 & 0.92 & 0.90 & 1.24 & 0.74\\ \hline
1L &  $12.46-16$ & $146839$ & $14.15$ & 0.05 & -0.08 & 11.3 & 11.5 & 8.75 & 2.26\\ 
2L &  $16-25$ & $182006$ & $20.25$ & 0.18 & -0.003 & 14.7 & 15.2 & 10.8 & 2.54\\
3L &  $25-40$ & $121146$ & $32.05$ & 0.40 & 0.12 & 20.5 & 21.5 & 14.0 & 2.96\\ 
4L &  $40-100$ & $114367$ & $65.99$ & 0.92 & 0.44 & 34.5 & 37.1 & 21.2 & 3.77\\ \hline
\end{tabular}
\caption{Mass range, number of haloes, average mass and bias parameters (both weighted by halo counts) for the nine different mass
bins used in our study. The first five belong to the small-box simulation; the last four to the large-box 
simulation. The bias parameters are computed according to the two models for the peak height described in the text: the spherical and 
ellipsoidal collapse models. All values have been obtained adopting a Gaussian filter.}
\label{tab:1}
\end{table*}

\begin{figure*}
\includegraphics[width = 3.0in,keepaspectratio=true]{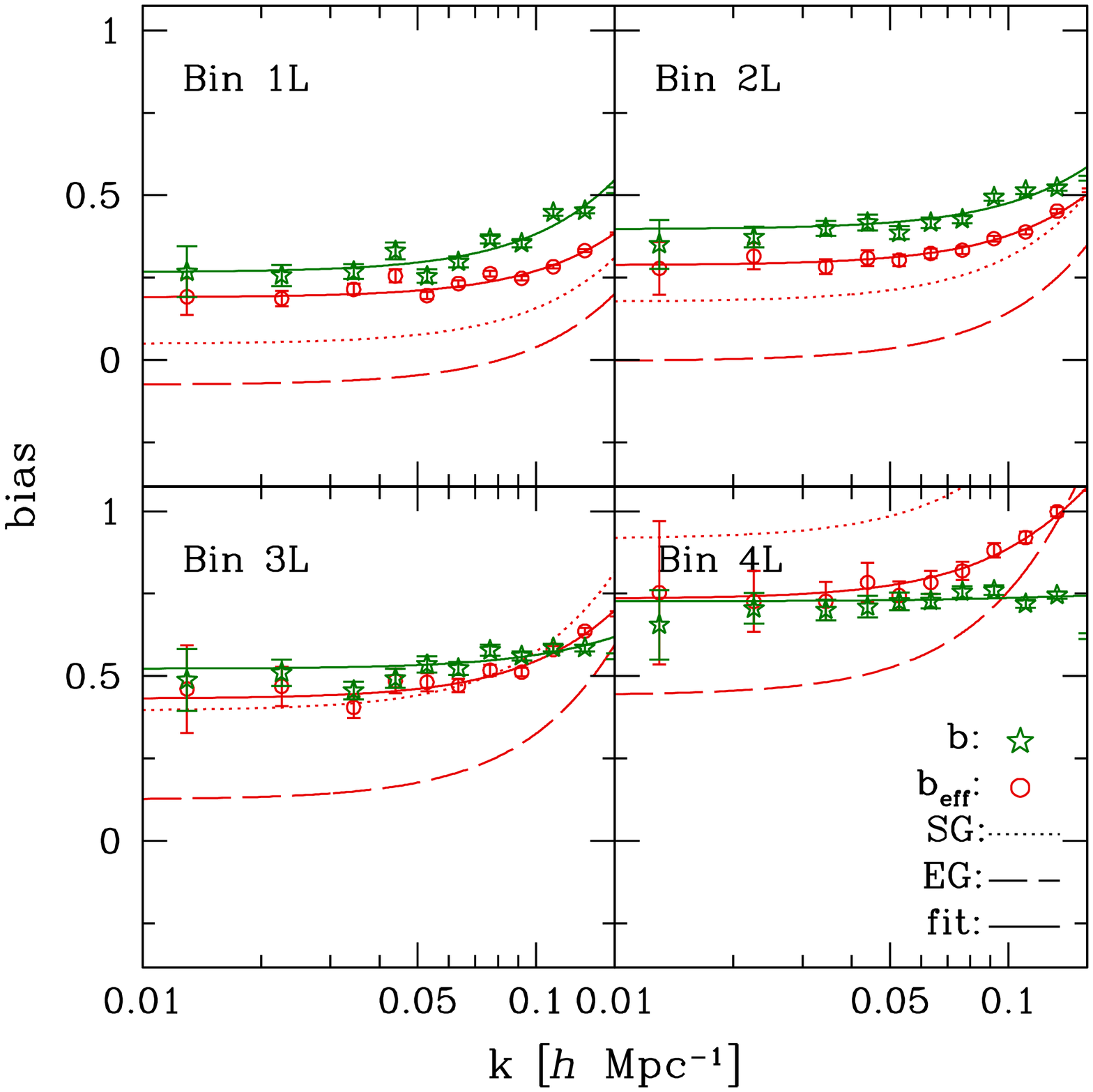}
\includegraphics[width = 3.0in,keepaspectratio=true]{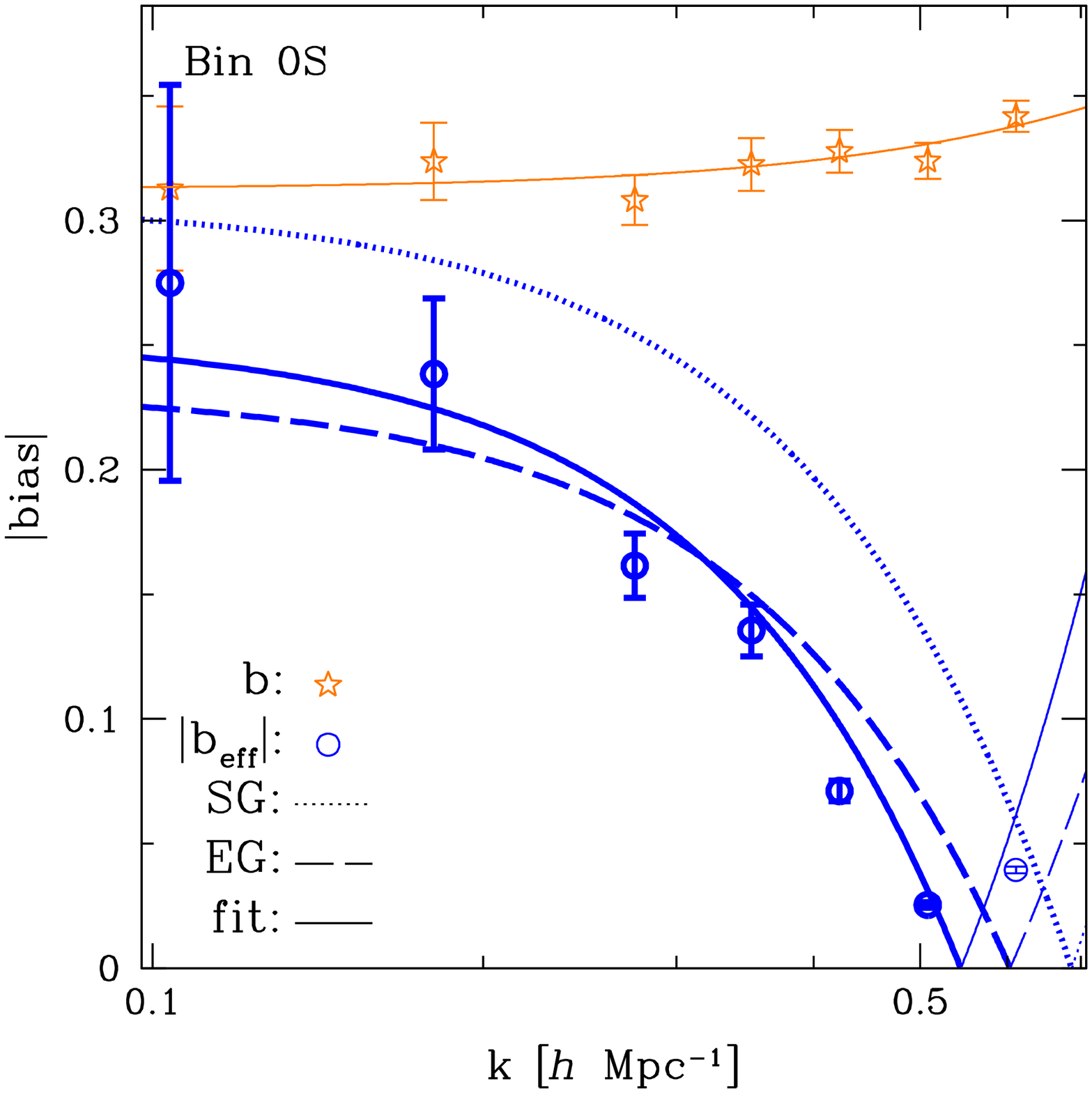}
\caption{
\textit{Left}: Stars and open circles plot, respectively, $b(k)$ and $b_{\rm eff}(k)$ for the
four separate mass bins in our large simulation box. Dotted lines show the predictions
obtained for the spherical collapse model, and dashed lines those of the ellipsoidal
collapse model adopting a Gaussian filter. Solid lines are the best-fits to the data
points. \textit{Right}: Stars and open circles correspond to $b(k)$ and $|b_{\rm eff}(k)|$, respectively.
As in the left panels, dotted and dashed line show to the SG and EG models, and 
solid lines are the best-fits to the data. Thick lines indicate negative values of the
bias. In all cases, errors are propagated assuming that the uncertainty in the 
power-spectrum is $\sigma(k)=\sqrt{2/N(k)}\cdot P(k)$, where $N(k)$ is the number of modes in 
each $k$ bin.}
\label{fig4}
\end{figure*}

It is clear that adopting a Gaussian filter entails lower values of $\delta_{\rm c}$ at all masses\footnote{At any mass 
scale, $M$, the smoothing length of a Gaussian filter exceeds that of a top-hat filter by a factor of 
$(3\sqrt{\pi /2})^{1/3}\approx 1.55$.}. Nonetheless, our results are still accurately described by eq. (\ref{o}), albeit with
slightly different values for the model parameters: $\delta_{*}=1.15$, $\alpha=0.37$ and $\beta=0.515$. We show this explicitly 
in Figure \ref{fig1} using a dashed line. The one-sigma dispersion about the mean trend is well described by a simple 
power-law, $\Sigma = 0.4 \ \sigma_0^{6/5}$, as seen in the upper panel of the plot. 

\section{Results}
\label{tre}

In this section we test the predictions of the DS model against the
measured density and velocity bias of dark matter protohaloes. In order to
do so, we calculate the (model-predicted) bias parameters $b_{\nu}$, $b_{\zeta}$ and $b_{\sigma}$
for each mass bin in both of our simulations. Since the values of $b_{\nu}$ and
$b_{\zeta}$ depend on peak height (and hence smoothing scale) we adopt two models
for the collapse barrier: one assumes the spherical collapse model
(hereafter SG) and the other the ellipsoidal collapse model (hereafter
EG). A Gaussian filter is used in both cases. Table 1 lists the bias
parameters predicted by the DS model for these choices of collapse
barrier.

\begin{table*}
\begin{tabular}{|c|c|c|c|c|c|c|c|c|}
\hline
Bin & \multicolumn{4}{|c|}{$b_{\nu}$} & \multicolumn{4}{|c|}{$b_{\zeta}$ ($h^{-2}\,\mathrm{Mpc}^2$)} \\ \hline
 &\multicolumn{2}{|c|}{$P_{\rm mh}$} & \multicolumn{2}{|c|}{$P_{\rm h}$} & \multicolumn{2}{|c|}{$P_{\rm mh}$} & \multicolumn{2}{|c|}{$P_{\rm h}$}\\ \hline
 &FOF &SO &FOF &SO &FOF &SO &FOF &SO \\ \hline
0S &  -0.25 & -0.24 & -0.31 & -0.30 & 0.86 & 0.81 & 0.14 & 0.11\\
1L &  0.19 & 0.22 & 0.27 & 0.37 & 8.8 & 7.9 & 12.6 & 12.0\\ 
2L &  0.29 & 0.31 & 0.40 & 0.39 & 10.3 & 9.3 & 9.5 & 8.8\\
3L &  0.43 & 0.47 & 0.52 & 0.54 & 13.9 & 13.1 & 6.7 & 5.9\\ 
4L &  0.73 & 0.75 & 0.73 & 0.73 & 21.8 & 20.9 & 6.4 & 5.4\\ \hline
\end{tabular}
\caption{Best-fit values for the density bias parameters from the cross-spectra and the auto-spectra. Uncertainties are always at the few per cent level. Haloes are identified either with the FOF or the SO algorithms.}
\label{tab:2}
\end{table*}

\begin{figure*}
\includegraphics[width = 3.0in,keepaspectratio=true]{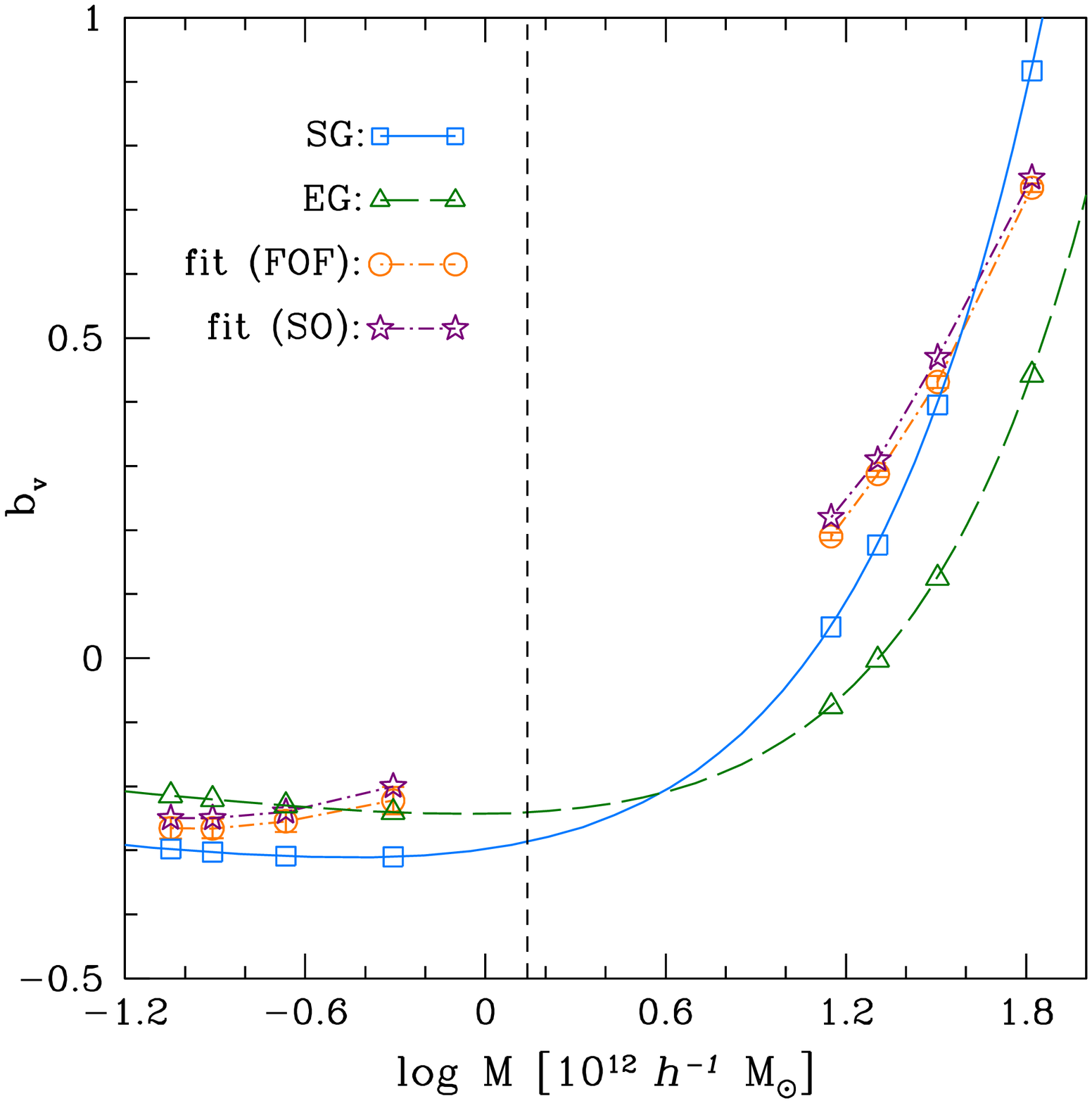}
\includegraphics[width = 3.0in,keepaspectratio=true]{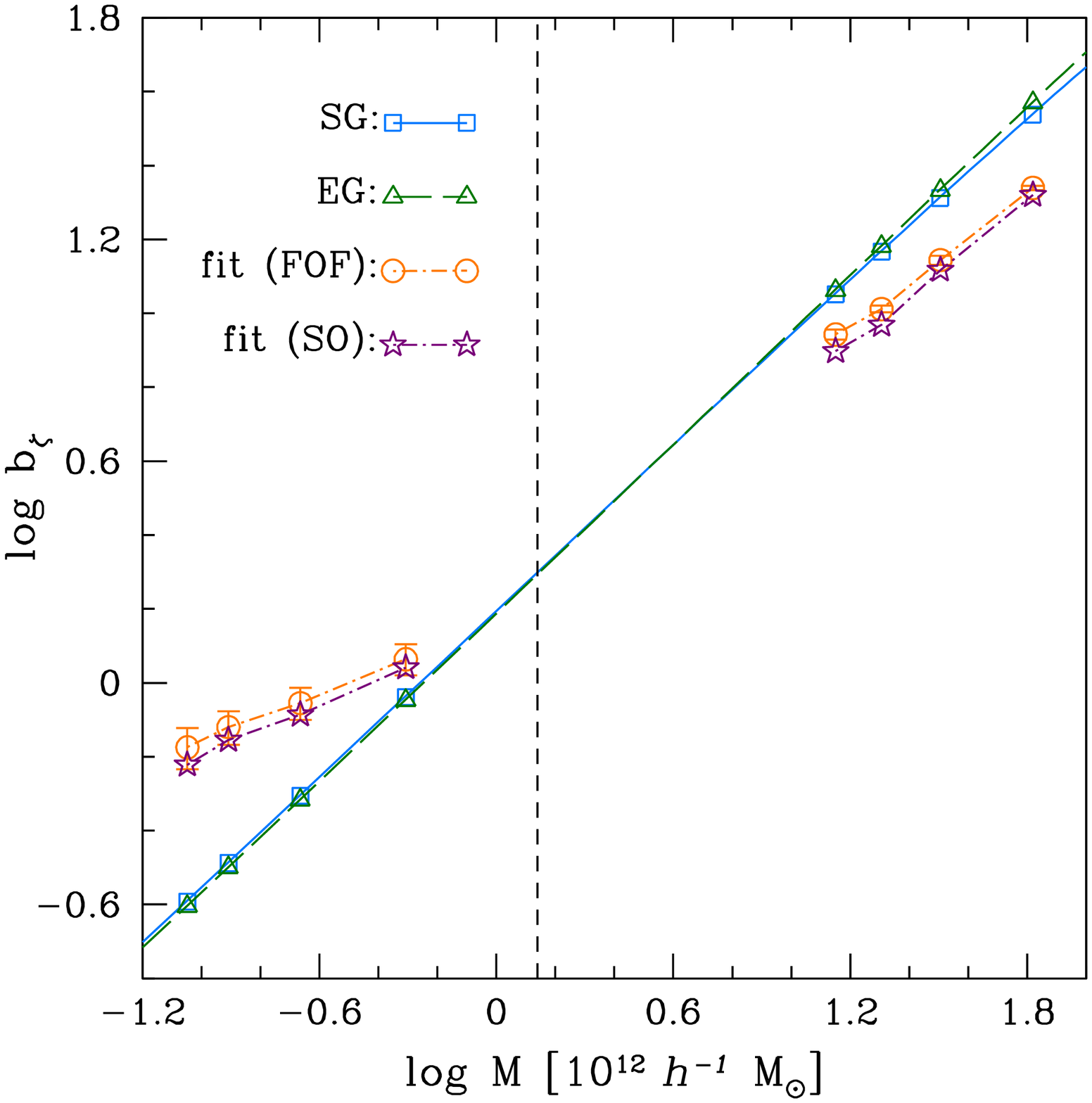}
\caption{
Density bias parameters, $b_{\nu}$ (\textit{left}) and $b_{\zeta}$ (\textit{right}), plotted as a function of halo mass.
Orange circles with error bars correspond to the best-fit values obtained from the 
cross-spectra of haloes identified with the FOF algorithm. The corresponding values for SO haloes are shown with purple stars. The solid and dashed lines show, respectively, the predictions of the DS model for the spherical collapse model, and for the ellipsoidal model with a Gaussian filter. The vertical dashed black lines mark $M_*$.}
\label{fig5}
\end{figure*}

\subsection{Density spectra}
In Figure \ref{fig4} we plot $b(k)$ and $b_{\mathrm{eff}}(k)$ extracted from our simulations,\footnote{The values directly measured from the simulations are rescaled by the growth factor $D(z_{\rm in})$ to match the theoretical estimates, where $\delta$ is linearly extrapolated to $z=0$. Note that the actual Lagrangian bias is a factor $D^{-1}\simeq 40-50$ larger than the values reported in the figure.} along with the model prediction (eq. (\ref{n})). 
Although the model is not expected to work for very small scales, we nonetheless show the results for each box up 
to $k\simeq 1/R_{\rm f}$ (we remind that $R_{\rm f}$ is the smoothing scale needed for the velocity field). 
Overall, the model expression for the initial peak bias is able to describe the simulation results reasonably well. This can be 
seen from the solid lines in Figure \ref{fig4}, where $b_{\nu}$ and $b_{\zeta}$ have been treated as free parameters and fitted 
to both the cross- \emph{and} the auto spectra. However, it is clear that the values for the fit parameters are different
in the two cases (see Table \ref{tab:2}), implying $b(k)\neq |b_{\mathrm{eff}}(k)|$ or, equivalently, $r(k) \neq 1$.
In particular, for the range of masses in Bin 0S, $b_{\mathrm{eff}}$ is negative, i.e. $r<0$. The predictions from
the SG and EG barriers, as apparent from the dashed and dotted lines in the right panel in Figure \ref{fig4}, are also negative. 
Hence the bias model matches more closely $b_{\mathrm{eff}}$ rather than $b$. This is completely expected, because eq. (\ref{n}) is exact only  
for the cross-correlation between peaks and matter while it neglects higher-order corrections for the peak autocorrelation.
However, neither the SG nor the EG barrier provide the appropriate values for 
the coefficients. In particular, the EG barrier performs better for low masses (bin 0S), while the situation is reversed 
for the higher mass bins.

Figure \ref{fig4} suggests that the stochasticity is more of a problem for haloes with $M<M_*$, and on smaller scales. Since high-mass haloes are highly correlated with density peaks in the initial conditions \citep{b8} and
peaks follow eq. (\ref{la}), the relationship between $\delta_{\rm h} ({\bf k})$ and $\delta({\bf k})$ is likely to be more deterministic for massive haloes.
Note, however, that the estimate of $r$ is affected by the shot-noise correction, which is large in our samples. Therefore we cannot draw definitive conclusions
regarding stochasticity.  

\begin{figure*}
\includegraphics[width = 3.0in,keepaspectratio=true]{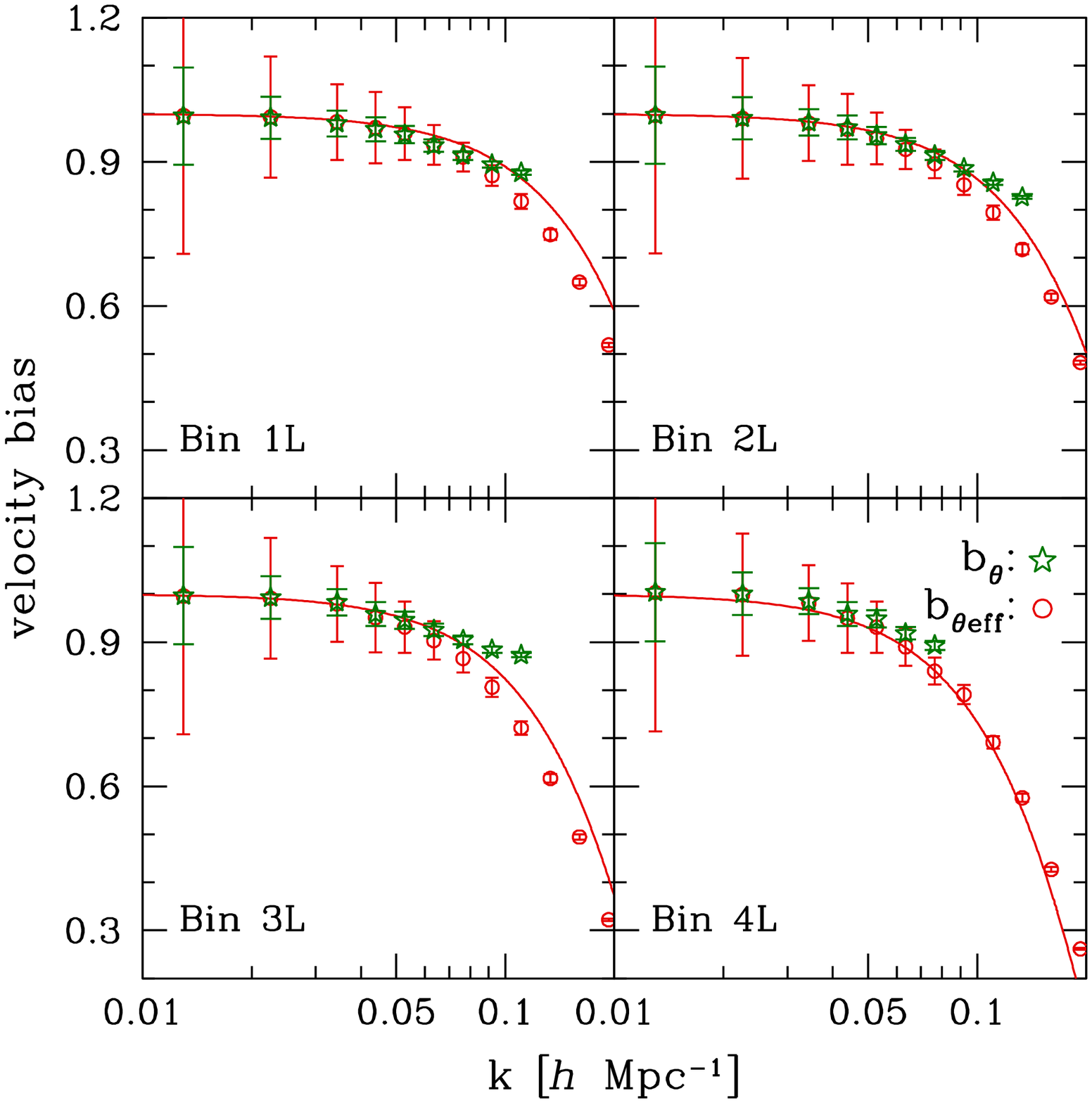}
\includegraphics[width = 3.0in,keepaspectratio=true]{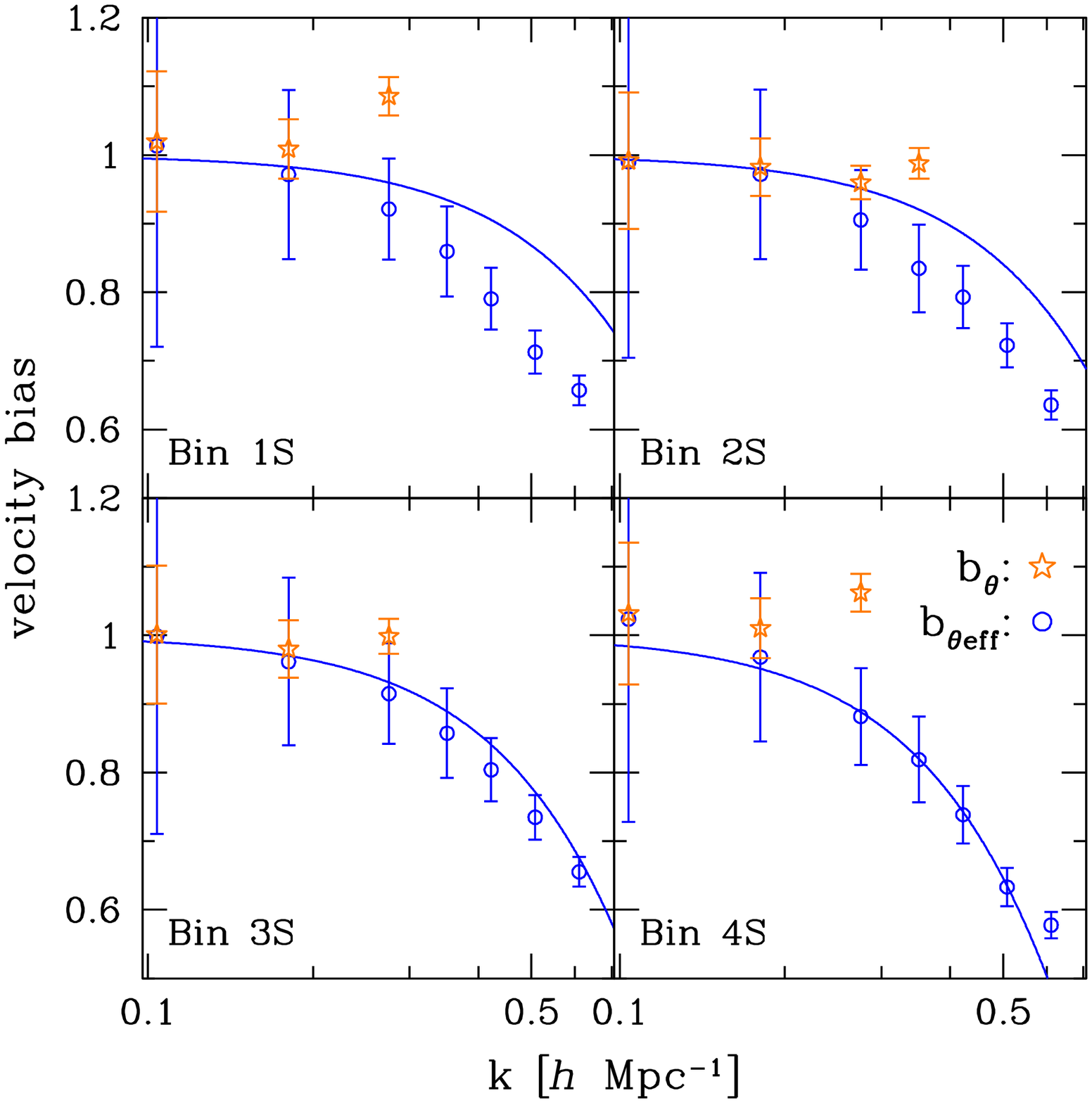}
\caption{
Stars and circles show $b_{\theta}(k)$ and $b_{\theta \mathrm{eff}}(k)$, respectively, for haloes in our large 
(\textit{left}) and small (\textit{right}) simulation boxes. The solid line is the DS model prediction (eq. (\ref{lb}) with $R_s$ given in Table \ref{tab:1}), which is independent of peak height. Errors are propagated as in Fig. \ref{fig4}. Note that we only show $b_{\theta}(k)$ on scales over which the shot-noise
is sub-dominant, as stated in Section \ref{analysis}.}
\label{fig6}
\end{figure*}

The performance of the model is characterized in another way in Figure \ref{fig5}, where we compare directly the mass dependence of the 
bias parameters $b_{\nu} (M)$ and $b_{\zeta} (M)$ predicted by the SG and EG models to the best-fit values obtained using the parameterization of the cross-spectrum in eq. (\ref{n}). We do not show the corresponding results from the auto-spectrum because of the uncertain shot-noise subtraction. Figure \ref{fig5} reveals that overall the $b_{\nu}$ values derived from the SG model are closer to the fitted ones. In contrast, there is almost 
no difference between the two models for $b_{\zeta}$. For both parameters the mass dependence follows a different trend when we 
compare the fits to the models. In particular, the prediction for $b_{\zeta}$ and $M<M_*$ becomes more and more inaccurate with decreasing halo mass (up to a factor of $\sim 3$ for $M\simeq M_*/15$). On the other hand, in the mass range for which the model was developed ($M>M_*$), $b_{\zeta}$ is overpredicted by a factor of $\sim 1.5$, independent of halo mass. 
Figure \ref{fig4} shows that the disagreement is even worse when $b_{\nu}$ and $b_{\zeta}$ are obtained from $P_{\rm h}$. All of this demonstrates that eqs. (\ref{ja}) and (\ref{jb}) cannot accurately describe the mass dependence of the Lagrangian density bias, especially for $M<M_*$. 
We will discuss possible reasons for this in Section \ref{sum}, but first turn our attention to the velocity spectra.

\subsection{Velocity spectra}

Figure \ref{fig6} shows $b_{\theta}(k)$ and $b_{\theta\mathrm{eff}}(k)$ for bins 1S-4S and 1L-4L; the theoretical model, $b_{\rm vel}(k)$, is also plotted. The data have been ``de-smoothed", i.e. they have been divided by a filter function with smoothing scale $R_{\rm f}$, the same used to smooth the velocity field in the first place.
It is apparent that on large scales the two estimates of the bias coincide, i.e. $r_{\theta}\simeq1$, indicating a strong correlation between the fields.
Overall, we can conclude that the velocity bias is deterministic to a good approximation.

There is an excellent agreement (better than $10$ per cent) between the simulation results
and the model predictions for all mass bins at scales $k<0.1 \ h\, \mathrm{Mpc}^{-1}$ for the large box and $k<0.3 \ h\, \mathrm{Mpc}^{-1}$ for the small one. Therefore, the peak model provides a faithful description of the velocity bias.

\section{Summary}
\label{sum}
We have investigated the Lagrangian bias of dark-matter haloes by testing the theoretical model proposed by DS against
N-body simulations of structure formation. The model assumes that haloes form from density peaks and predicts a scale-dependent bias for both the density and velocity fields. Our main results can be summarized as follows.
\begin{itemize}
\item When averaged over a spherical Lagrangian volume containing the appropriate mass, the linear density
contrast measured at protohalo centres depends sensitively on the choice of smoothing kernel. For 
a Gaussian kernel, for example, the resulting density contrasts are systematically lower than those
computed using a top-hat filter. This is because, at fixed mass, the smoothing length of a Gaussian kernel exceeds that of a
top-hat filter by a factor of about $1.55$ resulting in systematically lower density
estimates. Nonetheless, the median barrier height computed with a Gaussian kernel can be accurately parameterized 
by the same fitting formula first advocated by \citet{b9} for the case of a top-hat filter, albeit with different 
values for the numerical parameters. We use this result to approximate the collapse threshold for dark matter 
halo formation when adopting a Gaussian filter.

\item The functional forms for the density and velocity bias relations derived by DS - our eqs. (\ref{n}) and (\ref{nb}) -
accurately describe the results obtained from our simulations, 
provided the parameters of the model are allowed to vary with respect to
the model-predicted values.
In both cases, the Lagrangian bias is characterized by a constant term that dominates on very large scales, and a scale-dependent
term proportional to $k^2$.

\item Quantitatively, the velocity bias predicted by the DS model is
able to reproduce the measured protohalo velocity bias in our simulations 
to better than $10$ per cent, provided we limit ourselves to quasi-linear scales ($k\leq 0.3 \ h\, \mathrm{Mpc}^{-1}$).
The predicted density bias (eqs. (\ref{ja}) and (\ref{jb})), on the other hand, does not match the Lagrangian density bias extracted
from the simulations. This is likely due to the more complex nature of the density
bias, which depends additionally on peak height and on the exact definition of a protohalo. These results are
independent of whether one adopts a barrier height consistent with either the
spherical or ellipsoidal collapse model.

\item 
We have measured the mass dependence of the density bias, $b_{\nu}$ and $b_{\zeta}$, by fitting 
the density power-spectra obtained for haloes in several different mass bins in each
of our two simulations. The most massive haloes identified in our simulations are the
most strongly biased, and have characteristic overdensities that correlate with the
underlying matter density. In contrast, the distribution for low mass haloes 
($M_{\rm h} < 10^{12} h^{-1}\,M_{\odot}$) exhibits an 
anti-correlation. A similar trend is also predicted by the DS model, although with
noticeable differences (see Figure \ref{fig5}). We emphasize that the best-fit values for $b_{\nu}$ and $b_{\zeta}$
obtained in our analysis apply for a LCDM cosmogony (with our adopted cosmological 
parameters) and only over the limited mass range probed by our simulations. Future work 
should consider how the Lagrangian density bias depends on the underlying cosmology.
 
\item 
Our comparison of the auto- and cross-spectra in Section \ref{tre} suggests the presence of a
stochasticity in the Fourier-space density bias of dark matter protohaloes, but none for the
velocity bias. This is in disagreement with the DS model for the bias at leading order, and seems to corroborate the need for higher-order terms in the expression for the auto-spectrum of the protohaloes, as predicted in \citet{b47}. However, we cannot draw firm conclusions regarding stochasticity in 
the bias estimates due to uncertainties in the shot-noise subtraction.

\end{itemize}

We have tested the Lagrangian bias model of DS against a pair of high resolution N-body simulations
of structure formation and found that it is able to accurately reproduce the velocity bias, but not
that of the density. Our analysis focused on protohaloes, the high redshift progenitors of $z=0$ dark-matter haloes, whereas the model describes the biasing of density peaks. 
One possible explanation for the differences in the model predictions and simulation
results comes form the differences between the expressions for the bias parameters: $b_{\nu}$, $b_{\zeta}$ and $b_{\sigma}$.
The latter of the three is solely determined by the linear matter power spectrum and the smoothing 
scale corresponding to a given mass, $M$. The density bias parameters, however, depend additionally on 
explicit properties of the peaks, such as their height, $\nu$, mean curvature, $\bar{u}$, and on an assumed collapse
threshold for their identification as haloes of a given mass. Figure \ref{fig1} shows that there is a large 
halo-to-halo variation in $\nu$ at any given mass scale; characterizing the collapse threshold as a single
mass-dependent value may, therefore, be too simplistic. This added complexity introduces a significant 
margin for error in the model's estimates of the density bias. It remains to be seen whether more
realistic models of the collapse barrier - such as those that account for the statistical scatter in 
the linear over-densities of protohaloes at a given mass - will improve the model's predictive power.

Another possibility for the discrepancies stems from the fact that we are identifying linear density
peaks with protohaloes in our simulation initial conditions. Although the majority of our dark matter
protohaloes form in the vicinity of linear density peaks of the same characteristic mass, the fate of
{\em all} peaks with the same mass and overdensity is unclear. Uncertainties associated with the identification
of protohaloes within the linear density field may have
adverse effects on the predictive power of the density bias model. A more detailed understanding of how
protohaloes map onto linear density peaks, and vice versa, will, no doubt, provide valuable insight into 
the mechanisms behind halo biasing. 

\section*{Acknowledgments}
AE acknowledges financial support from the SFB-Transregio 33 ``The Dark Universe" by the Deutsche Forschungsgemeinschaft (DFG) and a scholarship from the Bonn-Cologne Graduate School (BCGS). We acknowledge interesting discussion on halo finders with Steffen Knollmann and Andrea Macci\`o. We would like to thank Jeremy Tinker for providing us with his SO halo finder, and for help applying it to our simulations. We are very grateful to the referee, Vincent Desjacques, for useful suggestions and valuable insights.

\label{lastpage}

\end{document}